\documentclass[a4paper,11pt]{article}

\usepackage{jheppub_mod} 
\usepackage[T1]{fontenc} 

\usepackage{amsmath,booktabs}
\usepackage{graphicx}
\usepackage{xspace}
\usepackage{color}
\pagecolor{white}
\usepackage[dvipsnames]{xcolor}
\usepackage{url}
\usepackage[utf8]{inputenc}
\usepackage{epstopdf}
\usepackage{hyperref}
\usepackage{multirow}

\newcommand{\mpi}{M_\pi}
\newcommand{\beq}{\begin{equation}}
\newcommand{\eeq}{\end{equation}}

\newcommand{\Li}{\text{Li}}

\newcommand{\GeV}{\,\text{GeV}}

\renewcommand{\Im}{\text{Im}\,}
\renewcommand{\Re}{\text{Re}\,}
\newcommand{\Ims}{\text{Im}_s\,}
\newcommand{\Imt}{\text{Im}_t\,}

\allowdisplaybreaks[1]

\title{Radiative corrections to the forward--backward asymmetry in $\boldsymbol{e^+e^-\to\pi^+\pi^-}$}

\author[a]{Gilberto Colangelo,}
\author[a]{Martin Hoferichter,}
\author[a]{Joachim Monnard,}
\author[a,b]{and Jacobo Ruiz de Elvira}

\affiliation[a]{Albert Einstein Center for Fundamental Physics, Institute for Theoretical Physics, University of Bern, Sidlerstrasse 5, 3012 Bern, Switzerland}

\affiliation[b]{Universidad Complutense de Madrid, Facultad de Ciencias F\'isicas,
Departamento de F\'isica Te\'ori-ca and IPARCOS, Plaza de las Ciencias 1, 28040 Madrid, Spain}

\emailAdd{gilberto@itp.unibe.ch}
\emailAdd{hoferichter@itp.unibe.ch}
\emailAdd{monnard@itp.unibe.ch}
\emailAdd{jacobore@ucm.es}

\abstract{We present a calculation of the $C$-odd radiative corrections to $e^+e^-\to \pi^+\pi^-$ in a dispersive formalism, concentrating on the leading pion-pole contribution in the virtual box diagrams. In particular, we show how the effect of a general pion vector form factor in the loop integral can be incorporated in a model-independent way and how the cancellation of infrared singularities proceeds in this case. The numerical results, dominated by the infrared enhanced contributions, indicate significant corrections beyond scalar QED, essentially confirming recent findings in generalized vector-meson-dominance models.}

\begin{document}

\maketitle
	
\section{Introduction}
\label{sec:intro}

\begin{sloppypar}
The data-driven determination of hadronic vacuum polarization (HVP) is dominated by the $2\pi$ channel, which gives about $70\%$ of the HVP contribution to the anomalous magnetic moment of the muon~\cite{Aoyama:2020ynm}. The consensus number~\cite{Davier:2017zfy,Keshavarzi:2018mgv,Colangelo:2018mtw,Hoferichter:2019gzf,Davier:2019can,Keshavarzi:2019abf}
\beq
a_\mu^\text{HVP, LO}\big|_{e^+e^-}=693.1(4.0)\times 10^{-10}
\label{HVPee}
\eeq
for this quantity relies on a set of measurements from SND~\cite{Achasov:2005rg,Achasov:2006vp}, CMD-2~\cite{Akhmetshin:2001ig,Akhmetshin:2003zn,Akhmetshin:2006wh,Akhmetshin:2006bx}, \mbox{BESIII~\cite{Ablikim:2015orh}}, and CLEO~\cite{Xiao:2017dqv} (and, when dispersive constraints are included~\cite{Colangelo:2018mtw,Ananthanarayan:2018nyx,Davier:2019can,Hoferichter:2019gzf,Stamen:2022uqh}, also on space-like data~\cite{Dally:1982zk,Amendolia:1986wj}), but is dominated by the precision data sets from 
BaBar~\cite{Aubert:2009ad,Lees:2012cj} and KLOE~\cite{Ambrosino:2008aa,Ambrosino:2010bv,Babusci:2012rp,Anastasi:2017eio}. 
Since Ref.~\cite{Aoyama:2020ynm} new data have become available from
SND~\cite{SND:2020nwa}, but resolving the tension between BaBar and KLOE requires new measurements at a similar level of precision, as are expected from CMD-3~\cite{Ryzhenenkov:2020vrk}, BaBar~\cite{Abbiendi:2022liz}, BESIII~\cite{BESIII:2020nme}, and Belle II~\cite{Belle-II:2018jsg}. 
This question has become increasingly urgent in view of recent lattice-QCD results~\cite{Borsanyi:2020mff,Ce:2022kxy,Alexandrou:2022amy} challenging the data-driven value~\eqref{HVPee} at least for the intermediate window~\cite{Blum:2018mom}, and critical for the interpretation~\cite{Crivellin:2020zul,Keshavarzi:2020bfy,Malaescu:2020zuc,Colangelo:2020lcg,Colangelo:2022vok} of the $4.2\sigma$ tension between experiment~\cite{Muong-2:2006rrc,Muong-2:2021ojo,Muong-2:2021ovs,Muong-2:2021xzz,Muong-2:2021vma} and Standard-Model theory~\cite{Aoyama:2020ynm,Aoyama:2012wk,Aoyama:2019ryr,Czarnecki:2002nt,Gnendiger:2013pva,Davier:2017zfy,Keshavarzi:2018mgv,Colangelo:2018mtw,Hoferichter:2019gzf,Davier:2019can,Keshavarzi:2019abf,Hoid:2020xjs,Kurz:2014wya,Melnikov:2003xd,Masjuan:2017tvw,Colangelo:2017qdm,Colangelo:2017fiz,Hoferichter:2018dmo,Hoferichter:2018kwz,Gerardin:2019vio,Bijnens:2019ghy,Colangelo:2019lpu,Colangelo:2019uex,Blum:2019ugy,Colangelo:2014qya}. 
\end{sloppypar}

A potential weak point in the analysis concerns the treatment of radiative corrections~\cite{Abbiendi:2022liz,WorkingGrouponRadiativeCorrections:2010bjp,Colangelo:2022jxc}, which are implemented in Monte-Carlo generators relying on scalar QED supplemented by the pion vector form factor (VFF) wherever possible~\cite{Campanario:2019mjh}, to capture the dominant corrections from the structure of the pion. In the case of final-state radiation (FSR), this approach does capture the dominant infrared (IR) enhanced effects~\cite{Hoefer:2001mx,Czyz:2004rj,Gluza:2002ui,Bystritskiy:2005ib}, but it is not guaranteed that corrections can be neglected in all kinematic configurations relevant for precision measurements of $e^+e^-\to \pi^+\pi^-$~\cite{JMPhDThesis}.

\begin{figure}[t]
	\centering
	\includegraphics[width=0.9\textwidth]{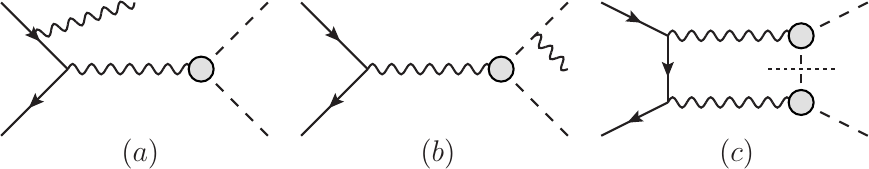}
	\caption{Representative diagrams contributing to $A_\text{FB}$ (solid, dashed, and wiggly lines denote electrons, pions, and photons, respectively). The interference of ISR $(a)$ and FSR $(b)$ produces real terms odd in $z$, while box diagrams $(c)$ give the virtual corrections. Only the sum of real and virtual contributions is IR finite. The gray blob denotes the pion VFF, which for $(b)$ can be justified because we only keep the soft limit. The short-dashed line in $(c)$ indicates that we only consider the pion-pole singularities, not general hadronic intermediate states that could contribute to the hadronic side of the box diagram.}
	\label{fig:diagrams}
\end{figure}

A possible test case concerns the forward--backward asymmetry in the same process, in which case radiative corrections arise from the interference of initial-state-radiation (ISR) and FSR diagrams~\cite{Binner:1999bt,Czyz:2004nq,BaBar:2015onb}
 combined with box diagrams, see Fig.~\ref{fig:diagrams}. While these corrections cancel when integrated over the entire phase space, they can be probed in the forward--backward asymmetry. Writing the Born cross section as
 \beq
 \label{Born}
 \frac{d\sigma_0}{dz}=\frac{\pi\alpha^2\beta^3}{4s}(1-z^2)\big|F_\pi^V(s)\big|^2,\qquad \beta=\sqrt{1-\frac{4\mpi^2}{s}},\qquad z=\cos\theta,\qquad \alpha=\frac{e^2}{4\pi},
 \eeq
 with the pion VFF $F_\pi^V(s)$ defined by the matrix element of the electromagnetic current $j_\text{em}^\mu=(2\bar u\gamma^\mu u-\bar d\gamma^\mu d-\bar s \gamma^\mu s)/3$,
 \beq
	\langle \pi^\pm(p') | j_\mathrm{em}^\mu(0) | \pi^\pm(p) \rangle =\pm (p'+p)^\mu F_\pi^V((p'-p)^2),
 \eeq
it is clear that the forward--backward asymmetry
\beq
A_\text{FB}(z)=\frac{\frac{d\sigma}{dz}(z)-\frac{d\sigma}{dz}(-z)}{\frac{d\sigma}{dz}(z)+\frac{d\sigma}{dz}(-z)}
\eeq
vanishes at tree-level, but radiative corrections give rise to odd terms in $z$. In Ref.~\cite{Arbuzov:2020foj} these corrections were calculated in scalar QED, multiplied in the end globally with $|F_\pi^V|^2$ to account for structure corrections.
However, in Ref.~\cite{Ignatov:2022iou} it was observed that this approximation was insufficient to describe preliminary data from the CMD-3 experiment, proposing an improved approach using generalized vector-meson-dominance (GVMD) models in the loop integral (see Ref.~\cite{Patil:1975ge} for a related analysis). 

In this paper, we present a model-independent approach that captures the effect of the pion-pole singularities in the virtual diagrams, and show that the dominant corrections still arise from the IR enhanced effects. For the real contribution, due to the interference of the ISR and FSR diagrams in Fig.~\ref{fig:diagrams}, we obtain the correction factor
\begin{align}
\label{real_final}
 \delta_\text{soft}&=\frac{2\alpha}{\pi}\Bigg\{\log\frac{\lambda^2}{4\Delta^2}\log\frac{1+\beta z}{1-\beta z}+\log(1-\beta^2)\log\frac{1+\beta z}{1-\beta z}+\log^2(1-\beta z)-\log^2(1+\beta z)\notag\\
 &+\Li_2\bigg(\frac{(z-1)\beta}{1-\beta}\bigg)+\Li_2\bigg(\frac{(1+z)\beta}{1+\beta}\bigg)
 -\Li_2\bigg(\frac{(z+1)\beta}{\beta-1}\bigg)-\Li_2\bigg(\frac{(1-z)\beta}{1+\beta}\bigg)\Bigg\},
\end{align}
which agrees with similar representations in Refs.~\cite{Arbuzov:2020foj,Ignatov:2022iou} and is defined relative to Eq.~\eqref{Born}
\beq
\frac{d\sigma}{dz}\bigg|_{C\text{-odd}}=\frac{d\sigma_0}{dz}\Big[\delta_\text{soft}(\lambda^2,\Delta)+\delta_\text{virt}\big(\lambda^2\big)\Big]+\frac{d\sigma}{dz}\bigg|_\text{hard}(\Delta).
\eeq
The parameter $\lambda=m_\gamma$, a small photon mass, regularizes the IR divergence and $\Delta$ is a cutoff in the photon energy (we do not consider the hard contribution with photon energy above $\Delta$ any further). As we will show below, already the cancellation  
of the IR singularity becomes quite subtle for a general pion VFF. After discussing the analytic structure in Sec.~\ref{sec:disp}, we derive our result for the virtual contribution in terms of the standard scalar loop functions in Sec.~\ref{sec:C0D0}, before turning to the numerical analysis in Sec.~\ref{sec:numerical}. Our conclusions are summarized in Sec.~\ref{sec:conclusions}.

\section{Dispersion relations and cut structure}
\label{sec:disp}

We define the kinematic variables according to
\beq
e^+(p_2)e^-(p_1)\to \pi^+(l_2)\pi^-(l_1),
\eeq
with Mandelstam variables
\begin{align}
 s&=(p_1+p_2)^2=(l_1+l_2)^2,\notag\\
 t&=(p_1-l_1)^2=(p_2-l_2)^2=\frac{1}{2}\Big(2\mpi^2+2m_e^2-s+z \sigma_e\sigma_\pi s\Big),\notag\\
 u&=(p_1-l_2)^2=(p_2-l_1)^2=\frac{1}{2}\Big(2\mpi^2+2m_e^2-s-z \sigma_e\sigma_\pi s\Big),
\end{align}
where
\beq
\sigma_\pi\equiv \beta=\sqrt{1-\frac{4\mpi^2}{s}},\qquad \sigma_e=\sqrt{1-\frac{4m_e^2}{s}},\qquad z=\frac{t-u}{s\sigma_\pi\sigma_e},
\eeq
and $s+t+u=2\mpi^2+2m_e^2$.
In the following, we will always put $m_e=0$ unless required to regularize collinear singularities.

To isolate the contribution from the pion pole in the hadronic part of the loop diagram, one starts from fixed-$s$ dispersion relations, which for scalar particles would immediately allow one to identify the scalar loop integral $D_0$~\cite{Denner:1991kt}  in terms of its double-spectral function. This procedure has been performed in detail for the pion boxes in hadronic light-by-light scattering~\cite{Colangelo:2015ama}, demonstrating that the non-box contributions that arise in a diagrammatic scalar-QED calculation are simply required by gauge invariance, in such a way that the final result can be obtained by multiplying the scalar-QED amplitude by the appropriate pion VFFs for the external photons.\footnote{One might wonder whether considering only the pion-pole contribution in the sub-amplitude $\gamma^*\gamma^*\to\pi\pi$ is a good approximation. To evaluate unitarity corrections to the pion-pole contribution one could start from a fixed-$t$ instead of a fixed-$s$ dispersion relation and make use of the full amplitudes for $\gamma^*\gamma^*\to\pi\pi$~\cite{Garcia-Martin:2010kyn,Hoferichter:2011wk,Moussallam:2013una,Danilkin:2018qfn,Hoferichter:2019nlq,Danilkin:2019opj}. However, as shown by these studies, below $1\GeV$ the Born terms yield the dominant contribution.}

In the present case a similar argument applies if a dispersive representation is used for the pion VFF in the $e^+e^-\to \pi^+\pi^-$ subamplitudes, i.e., if we replace
\beq
\label{VFF}
\frac{F_\pi^V(s)}{s}=\frac{1}{s}+\frac{1}{\pi}\int_{4\mpi^2}^\infty d s'\frac{\Im F_\pi^V(s')}{s'(s'-s)}\to \frac{1}{s-\lambda^2}-\frac{1}{\pi}\int_{4\mpi^2}^\infty d s'\frac{\Im F_\pi^V(s')}{s'} \frac{1}{s-s'}
\eeq
for each of the two photon propagators. In Eq.~\eqref{VFF} we have introduced the IR regulator in the pole terms, which are the ones that produce the IR divergence. 
The resulting representation for the box diagram then consists of pole times pole, mixed pole and dispersive, and dispersive times dispersive contributions. In the next section, we will express each of them in terms of standard loop functions.

\section{Formulation in terms of scalar loop functions}
\label{sec:C0D0}

The tensor decomposition is easiest to derive directly for the spin sum of the interference of the box diagram with the tree-level amplitude. This gives the decomposition of the virtual correction
\begin{align}
\delta_\text{virt}&=\bar \delta_\text{virt}\big(\lambda^2,\lambda^2\big)-\frac{1}{\pi}\int_{4\mpi^2}^\infty ds'\frac{\Im F_\pi^V(s')}{s'} \big[\bar \delta_\text{virt}\big(s',\lambda^2\big)+\bar \delta_\text{virt}\big(\lambda^2,s'\big)\big]\notag\\
&+\frac{1}{\pi}\int_{4\mpi^2}^\infty ds'\frac{\Im F_\pi^V(s')}{s'} \frac{1}{\pi}\int_{4\mpi^2}^\infty ds''\frac{\Im F_\pi^V(s'')}{s''}\bar\delta_\text{virt}(s',s''),
\end{align}
with 
\begin{align}
\label{tensor_decomposition}
 \bar\delta_\text{virt}(s',s'')&=-\frac{\Re F_\pi^V(s)}{2\beta^2 s(1-z^2)|F_\pi^V(s)|^2}\frac{\alpha}{\pi}\notag\\
 &\times\Re\bigg[
 4t\big(\mpi^2-t\big)\Big(C_0\big(m_e^2,t,\mpi^2,s',m_e^2,\mpi^2\big)+C_0\big(m_e^2,t,\mpi^2,s'',m_e^2,\mpi^2\big)\Big)
 \notag\\
 &\quad-4u\big(\mpi^2-u\big)\Big(C_0\big(m_e^2,u,\mpi^2,s',m_e^2,\mpi^2\big)+C_0\big(m_e^2,u,\mpi^2,s'',m_e^2,\mpi^2\big)\Big)
 \notag\\
 &\quad-4s(t-u)C_0\big(m_e^2,s,m_e^2,m_e^2,s',s''\big)
 +4\big(t^2-u^2\big)C_0\big(\mpi^2,s,\mpi^2,\mpi^2,s',s''\big)\notag\\
 &\quad+4\big(\mpi^2-t\big)\Big(\big(\mpi^2-t)^2+\mpi^4+t(s'+s''-u)\Big)\notag\\
 &\quad\qquad\times D_0\big(m_e^2,m_e^2,\mpi^2,\mpi^2,s,t,s',m_e^2,s'',\mpi^2\big)\notag\\
 &\quad-4\big(\mpi^2-u\big)\Big(\big(\mpi^2-u)^2+\mpi^4+u(s'+s''-t) \Big)\notag\\
 &\quad\qquad\times D_0\big(m_e^2,m_e^2,\mpi^2,\mpi^2,s,u,s',m_e^2,s'',\mpi^2\big)\bigg] + (\Re\to \Im),
\end{align}
and loop functions $C_0$, $D_0$ in the conventions of Ref.~\cite{Denner:1991kt}.  We stress that the presence of $|F_\pi^V(s)|^2$ in the denominator is just an effect of the normalization of the corrections $\delta$ to the Born cross section~\eqref{Born}, whereas $F_\pi^V(s)$ in the numerator is due to the interference with the Born diagram.

\subsection{Pole--pole}

We first consider the case $s'=s''=\lambda^2$. The IR divergences are easiest to extract using dispersive representations of the loop integrals, leading to the expressions given in App.~\ref{sec:loop_pole_pole}.
 In combination, we find the following expression for the pole--pole contribution
\begin{align}
\label{pole_pole}
  \delta_\text{virt}^\text{pole--pole}&=\frac{\alpha}{\pi}\frac{\Re F_\pi^V(s)}{|F_\pi^V(s)|^2}\Bigg\{2\log\frac{\lambda^2}{s}\log\frac{1-\beta z}{1+\beta z} +\frac{z}{(1-z^2)\beta}\bigg[\frac{(1-\beta)^2}{\beta}\frac{\pi^2}{6}+2\log^2 2\notag\\
  &\quad-\log^2(1-\beta^2)
  +\frac{1+\beta^2}{2\beta}\bigg(4\Li_2\bigg(\frac{\beta-1}{1+\beta}\bigg)+\log^2\frac{1-\beta}{1+\beta}\bigg)\bigg]\notag\\
  &\quad+\frac{1+2\beta z+\beta^2}{(1-z^2)\beta^2}\bigg[\frac{1}{2}\log^2(1+\beta z)+\log\beta_+\log\frac{1+2\beta z+\beta^2}{1+\beta z}+\Li_2\big(\beta_+\big)\bigg]\notag\\
  &\quad-\frac{1-2\beta z+\beta^2}{(1-z^2)\beta^2}\bigg[\frac{1}{2}\log^2(1-\beta z)+\log\beta_-\log\frac{1-2\beta z+\beta^2}{1-\beta z}+\Li_2\big(\beta_-\big)\bigg]\Bigg\}\notag\\
  &+\frac{\alpha}{\pi}\frac{\Im F_\pi^V(s)}{|F_\pi^V(s)|^2}\frac{\pi}{(1-z^2)\beta^2}\bigg[(1+\beta^2)z\log\frac{1-\beta}{1+\beta}+2\beta z\log\frac{1-\beta^2 z^2}{1-\beta^2}\notag\\
  &\quad-\Big(1+\beta^2(2z^2-1)\Big)\log\frac{1-\beta z}{1+\beta z}\bigg],\qquad \beta_\pm = \frac{1-\beta^2}{2(1\pm\beta z)}.
\end{align}
The functional form of the contribution proportional to $\Re F_\pi^V/|F_\pi^V|^2$ agrees with Ref.~\cite{Arbuzov:2020foj}, i.e.,
as expected, the point-like limit is recovered from the pole--pole contribution upon setting  $F_\pi^V=1$.
In particular, we confirm that all collinear singularities cancel. 

\subsection{Pole--dispersive}

The loop integrals required for the mixed pole and dispersive contributions are provided in  App.~\ref{sec:loop_pole_dispersive}. 
As a first step, we consider the IR divergence: for the real part, we find
\begin{align}
\label{pole_disp_IR}
\delta_\text{virt}^\text{pole--disp}\big|_\text{Re}^\text{IR}&=\frac{\alpha}{\pi}\frac{\Re F_\pi^V(s)}{|F_\pi^V(s)|^2} \bigg[2\log\frac{\lambda^2}{s}\log\frac{1-\beta z}{1+\beta z} \Re \frac{s}{\pi}\int_{4\mpi^2}^\infty ds'\frac{\Im F_\pi^V(s')}{s'(s'-s)}\bigg]\notag\\
&=\frac{\alpha}{\pi}\frac{\Re F_\pi^V(s)}{|F_\pi^V(s)|^2} \bigg[2\log\frac{\lambda^2}{s}\log\frac{1-\beta z}{1+\beta z} \big(\Re F_\pi^V(s) -1\big)\bigg],
\end{align}
which, together with Eq.~\eqref{pole_pole}, combines to a prefactor $(\Re F_\pi^V)^2/|F_\pi^V|^2$. The IR divergence in the imaginary part is more subtle. It arises from the imaginary part generated by the $D_0$ function for $s'<s$, since the integration over $s'$ displays an end-point singularity at $s'=s$. To extract this singularity, we write 
\begin{align}
\label{IR_Im}
\frac{1}{\pi}\int_{4\mpi^2}^{s}ds' \frac{\Im F_\pi^V(s')}{s'} \Im \delta^\text{pole--disp}(s')&= 
\frac{1}{\pi}\int_{4\mpi^2}^{s}ds' \frac{\Im F_\pi^V(s')-\Im F_\pi^V(s)}{s'} \Im \delta^\text{pole--disp}(s')\notag\\
&+\frac{\Im F_\pi^V(s)}{\pi}\int_{4\mpi^2}^{s}ds' \frac{\Im \delta^\text{pole--disp}(s')}{s'}, 
\end{align}
with integrand 
\begin{align}
\Im \delta^\text{pole--disp}(s')&=\frac{2\pi\big(s'-s+(s+s'-2s z^2)\beta^2\big)\log\frac{1-\beta z}{1+\beta z}}{(s'-s)(1-z^2)\beta^2}\notag\\
&-\frac{2\pi z\Big((1+\beta^2)\log\frac{1-\beta}{1+\beta}+2\beta\log\frac{1-\beta^2z^2}{1-\beta^2}\Big)}{(1-z^2)\beta^2}.
\end{align}
The second integral in Eq.~\eqref{IR_Im}, however,
needs to be evaluated including the effect of the IR regulator, which becomes easier in dimensional regularization, e.g., by introducing another regulator $\delta$ 
\beq
\label{regulator}
\frac{1}{s'-s}\to \frac{1}{s'-s-i\delta}
\eeq
for the endpoint divergence in the Cauchy kernel~\cite{CottiniHolz}. The result reads 
\begin{align}
\label{pole_disp_Im}
 \delta_\text{virt}^\text{pole--disp}\big|_\text{Im}&=
 \frac{\alpha}{\pi}\frac{\Im F_\pi^V(s)}{|F_\pi^V(s)|^2}\bigg[
 \frac{1}{\pi}\int_{4\mpi^2}^s ds'\frac{\Im F_\pi^V(s')-\Im F_\pi^V(s)}{s'} \Im \delta^\text{pole--disp}(s')\bigg]\\
 &+\frac{2\alpha}{\pi}\frac{\big(\Im F_\pi^V(s)\big)^2}{|F_\pi^V(s)|^2}
\Bigg\{\log\frac{1-\beta z}{1+\beta z}\bigg(\log\frac{\lambda^2}{s}+2\log\frac{1-\beta^2}{\beta^2}\bigg)\notag\\
&\quad+\frac{\log(1-\beta^2)}{(1-z^2)\beta^2}\notag\\
&\quad\quad\times\bigg[2\beta z\log\frac{1-\beta^2z^2}{1-\beta^2}+z\big(1+\beta^2\big)\log\frac{1-\beta}{1+\beta}-\big(1+\beta^2z^2\big)\log\frac{1-\beta z}{1+\beta z}\bigg]\notag\\
&\quad+\log^2(1+\beta z)-\log^2(1-\beta z)\notag\\
 &\quad-\Li_2\bigg(\frac{(z-1)\beta}{1-\beta}\bigg)-\Li_2\bigg(\frac{(1+z)\beta}{1+\beta}\bigg)
 +\Li_2\bigg(\frac{(z+1)\beta}{\beta-1}\bigg)+\Li_2\bigg(\frac{(1-z)\beta}{1+\beta}\bigg)\Bigg\}.\notag
\end{align}
This expression decomposes into a part obtained by only keeping the photon-mass regulator $\lambda^2$ in the integration boundary plus a correction proportional to the finite terms of $\delta_\text{soft}$ in Eq.~\eqref{real_final}. The same result can also be obtained entirely using the photon-mass regulator, by expressing the additional integrals in terms of Eikonals~\cite{Budassi:2024whw}. 

Summing up the IR divergences in Eqs.~\eqref{pole_pole}, \eqref{pole_disp_IR}, and~\eqref{pole_disp_Im}, we finally obtain
\beq
\delta_\text{virt}^\text{IR}=2\frac{\alpha}{\pi} \log\frac{\lambda^2}{s}\log\frac{1-\beta z}{1+\beta z},
\eeq
which cancels against the real emission~\eqref{real_final}, as expected. However, we stress that for the virtual contribution the factor $|F_\pi^V|^2$ originates from a subtle interplay of three different contributions, which no longer works for the finite terms: for the latter $F_\pi^V$ remains buried inside dispersive integrals and neither factorizes nor combines into just the modulus squared. Ultimately, this is the reason why multiplying the point-like result by $|F_\pi^V|^2$ is a poor approximation.  

For completeness, we also give the full expression for the real part:
\begin{align}
\label{pole_disp_Re}
 \delta_\text{virt}^\text{pole--disp}\big|_\text{Re}&=
 \frac{\alpha}{\pi}\frac{\Re F_\pi^V(s)}{|F_\pi^V(s)|^2}
 \frac{1}{\pi}\int_{4\mpi^2}^\infty ds'\frac{\Im F_\pi^V(s')}{s'} \Re \delta^\text{pole--disp}(s'),\notag\\
 \Re \delta^\text{pole--disp}&=\frac{2s}{s'-s}\bigg[\log\frac{\lambda^2 s'}{(s'-s)^2}\log\frac{1-\beta z}{1+\beta z}+\Li_2\bigg(1-\frac{s'(1-\beta')}{s(1-\beta z)}\bigg)\notag\\
 &\quad+\Li_2\bigg(1-\frac{s'(1+\beta')}{s(1-\beta z)}\bigg)-\Li_2\bigg(1-\frac{s'(1-\beta')}{s(1+\beta z)}\bigg)-\Li_2\bigg(1-\frac{s'(1+\beta')}{s(1+\beta z)}\bigg)\bigg]\notag\\
 &+\frac{2(1+\beta^2)}{(1-z^2)\beta^2}\log\frac{1-\beta z}{1+\beta z}\log\frac{2s}{(1-\beta^2)|s-s'|}\notag\\
 &-\frac{4z}{(1-z^2)\beta}\bigg[s \Re \bar C_0\big(s,s',m_e^2\big)
 +\frac{s}{2}(1+\beta^2) \Re C_0\big(s,s',\mpi^2\big)+\log^2 2\notag\\
 &\quad+\log^2(1-\beta')+\log^2(1+\beta')-\log^2(1-\beta^2)-\log\frac{1-\beta^2}{2}\log\big(1-\beta^2z^2\big)\notag\\
 &\quad+\log\frac{1-\beta^2 z^2}{(1-\beta^2)^2}\log\frac{s}{s'}-\log\frac{s'(1-\beta^2)}{s(1-\beta^2 z^2)}\log\frac{s'}{|s-s']}\bigg]\notag\\
 &+\frac{2(1-2\beta z+\beta^2)}{(1-z^2)\beta^2}\bigg[\Li_2\bigg(1-\frac{s'(1-\beta')}{s(1-\beta z)}\bigg)+\Li_2\bigg(1-\frac{s'(1+\beta')}{s(1-\beta z)}\bigg)+\Li_2\big(\beta_-\big)\notag\\
 &\quad+\log\big(1-2\beta z +\beta^2\big)\log\beta_-+\frac{3}{2}\log^2(1-\beta z)\bigg]\notag\\
 &-\frac{2(1+2\beta z+\beta^2)}{(1-z^2)\beta^2}\bigg[\Li_2\bigg(1-\frac{s'(1-\beta')}{s(1+\beta z)}\bigg)+\Li_2\bigg(1-\frac{s'(1+\beta')}{s(1+\beta z)}\bigg)+\Li_2\big(\beta_+\big)\notag\\
 &\quad+\log\big(1+2\beta z +\beta^2\big)\log\beta_++\frac{3}{2}\log^2(1+\beta z)\bigg],
\end{align}
where 
\beq
\bar C_0\big(s,s',m_e^2)=\bigg[C_0\big(s,s',m_e^2)+\frac{\log\frac{m_e^2}{s'}\log\frac{|s'-s|}{s'}}{s}\bigg]_{m_e\to 0}
\eeq
is the finite part of the massless $C_0(s,s',m_e^2)$ loop function and 
again all the collinear singularities from $m_e\to 0$ cancel. The singularity at $s'=s$ is to be interpreted as the principal value, but in contrast to the imaginary part no end-point singularity arises. However, the imaginary part in Eq.~\eqref{regulator}, in combination with $\Im\delta^\text{pole--disp}(s')$ for $s'<s$, then produces a mixed correction that can interfere with $\Re F_\pi^V(s)$~\cite{Ignatov}
\beq
\label{delta_mixed}
\delta_\text{virt}^\text{pole--disp}\big|_\text{mixed}=-\frac{\alpha}{\pi}\frac{\Re F_\pi^V(s)\Im F_\pi^V(s)}{|F_\pi^V(s)|^2}2\pi\log\frac{1-\beta z}{1+\beta z}.
\eeq

\subsection{Dispersive--dispersive}

The purely dispersive correction can be expressed as
\begin{align}
\label{disp_disp}
 \delta_\text{virt}^\text{disp--disp}&=
 \frac{\alpha}{\pi}\frac{\Re F_\pi^V(s)}{|F_\pi^V(s)|^2}
 \frac{1}{\pi}\int_{4\mpi^2}^\infty ds'\frac{\Im F_\pi^V(s')}{s'} 
 \frac{1}{\pi}\int_{4\mpi^2}^\infty ds''\frac{\Im F_\pi^V(s'')}{s''}
 \Re \delta^\text{disp--disp}(s',s'')\notag\\
 &+\frac{\alpha}{\pi}\frac{\Im F_\pi^V(s)}{|F_\pi^V(s)|^2}
 \frac{1}{\pi}\int_{4\mpi^2}^\infty ds'\frac{\Im F_\pi^V(s')}{s'} 
 \frac{1}{\pi}\int_{4\mpi^2}^\infty ds''\frac{\Im F_\pi^V(s'')}{s''}
 \Im \delta^\text{disp--disp}(s',s''),
\end{align}
where the integrands follow directly from Eq.~\eqref{tensor_decomposition}. All loop functions that contribute in this case, explicit expressions for which are provided in App.~\ref{sec:loop_dispersive_dispersive},  are IR finite and free of collinear singularities.  

\section{Numerical analysis}
\label{sec:numerical}

For the numerical analysis we use the pion VFF from Ref.~\cite{Colangelo:2018mtw} for $s\leq s_\text{cut}$, $s_\text{cut}=1\GeV^2$. 
Since the corrections $\delta$ are defined relative to the tree-level result~\eqref{Born}, which itself depends on $F_\pi^V$,
they need to be determined in a self-consistent way. Accordingly, we restrict the analysis here to the energy below $s_\text{cut}$. However, for the dispersive integrals we also need to provide $\Im F_\pi^V$ above, in particular, in writing Eq.~\eqref{VFF} we have implicitly assumed the sum rule
\beq
\frac{1}{\pi}\int_{4\mpi^2}^\infty ds' \frac{\Im F_\pi^V(s')}{s'}=1.
\eeq
In practice, this sum rule is easiest to fulfill by including excited $\rho'$, $\rho''$ resonances in the $\pi\pi$ phase shift, which is one of the variants for the asymptotic continuation studied in Ref.~\cite{Colangelo:2018mtw} (using the implementation from Ref.~\cite{Schneider:2012ez}, based on the data from Ref.~\cite{Belle:2008xpe}). In the following, we use the corresponding input for $F_\pi^V$. 

\begin{figure}[t]
	\centering
	\includegraphics[width=0.49\textwidth]{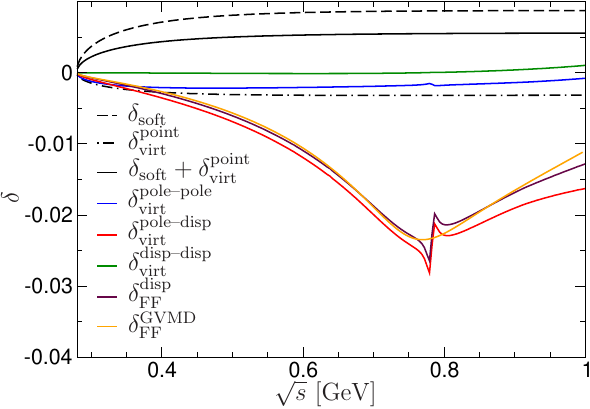}
	\includegraphics[width=0.49\textwidth]{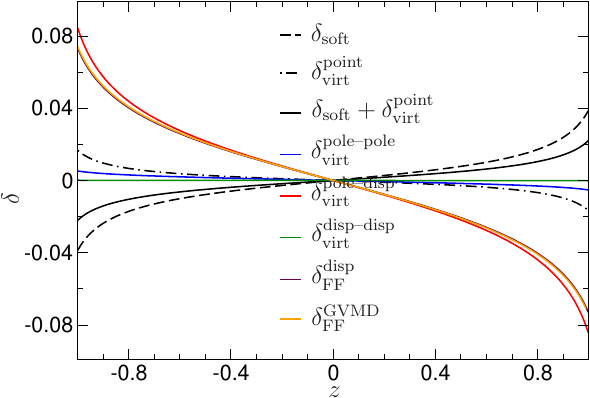}
	\caption{Correction factors $\delta$ as a function of $\sqrt{s}$ for fixed $z=\cos(1)$ (left) and as a function of $z$ for fixed $\sqrt{s}=0.75\GeV$ (right). The black lines denote the point-like result (dashed: real, dot-dashed: virtual, solid: sum of real and virtual), where in all cases the logarithmic terms including the IR divergence are not shown. In the same convention, we show our results for the pole--pole (blue), pole--dispersive (red), and dispersive--dispersive (green) contributions, as well as their sum minus the point-like virtual correction (maroon). The same quantity in the GVMD model of Ref.~\cite{Ignatov:2022iou} with a single Breit--Wigner is given for comparison (orange).}
	\label{fig:numerics}
\end{figure}

\begin{figure}[t]
	\centering
	\includegraphics[width=0.6\textwidth,angle=-90]{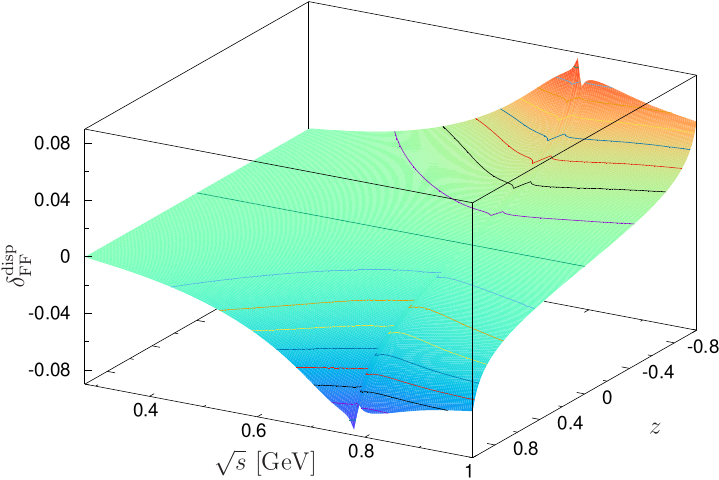}
	\caption{Structure-dependent correction $\delta_\text{FF}$ in the dispersive approach (sum of pole--pole, pole--dispersive, and dispersive--dispersive minus point-like virtual), as a function of $\sqrt{s}$ and $z$. The contour lines give increments of $0.01$. A data file is attached as supplemental material.}
	\label{fig:numerics_2d}
\end{figure}

Our numerical results for the corrections $\delta$ are shown in Figs.~\ref{fig:numerics} and~\ref{fig:numerics_2d}, where in all cases the terms proportional to $\log\frac{\lambda^2}{4\Delta^2}$ and $\log\frac{\lambda^2}{s}$, respectively, are dropped. In addition to the separate curves for the pole--pole, pole--dispersive, and dispersive--dispersive contributions, we also show the quantity $\delta_\text{FF}$, which was defined in Ref.~\cite{Ignatov:2022iou} as the total minus the point-like virtual correction. We also include the GVMD model from the same reference (in the variant using a single Breit--Wigner function for the $\rho(770)$), and, in Fig.~\ref{fig:numerics},  choose the same fixed parameters ($z=\cos(1)$ and $\sqrt{s}=0.75\GeV$, respectively) to facilitate the comparison. Figure~\ref{fig:numerics_2d} shows our results for $\delta_\text{FF}$ as a function of both $\sqrt{s}$ and $z$.

The main observation is that we confirm significant departures from the point-like approximation, which, as remarked in the previous section, are a remnant of the intricate manner how IR singularities cancel in the sum of real and virtual contributions for a general VFF. In particular, while the pole--pole piece remains small and actually stays close to the point-like result, the pole--dispersive correction receives a large enhancement in the vicinity of the $\rho$ resonance. The result is remarkably close to the GVMD approximation, much better than could be expected in view of its non-analytic nature. Finally, we find that the dispersive--dispersive contributions are negligible below $1\GeV$, supporting the expectation that the most important radiative corrections are the ones that display some form of IR enhancement.

\section{Conclusions}
\label{sec:conclusions}

In this paper we presented a calculation of the radiative corrections to the forward--backward asymmetry in $e^+e^-\to\pi^+\pi^-$ that includes the full effect of the pion vector form factor in the loop integral by means of a dispersive representation and thus 
captures the leading hadronic intermediate state. In particular, we studied how the cancellation of infrared divergences proceeds in the case of a general form factor, and found that an intricate interplay between a point-like pion-pole contribution and the real and imaginary parts of a mixed pole and dispersive correction becomes necessary. Overall, the numerical results support recent findings in a generalized vector-meson-dominance model~\cite{Ignatov:2022iou}, indicating significant deviations from the point-like approximation, but the dispersive analysis puts this conclusion on a more solid foundation and allows one to trace back the origin of the large correction to a remnant of the infrared singularities. This implies that the usual assumption that the most important radiative corrections for processes involving hadrons should be the ones that display some form of infrared enhancement actually proves correct, while the problem in the scalar-QED calculation multiplied by the pion vector form factor was that these pieces were not correctly identified. These insights should prove valuable for reassessing the role of radiative corrections in precision measurements of $e^+e^-\to\pi^+\pi^-$.

\acknowledgments   
We thank Peter Stoffer for comments on the manuscript. We further thank the authors of Ref.~\cite{Budassi:2024whw} for alerting us to the scale ambiguity in our previous treatment of the endpoint singularity, Fedor Ignatov for pointing out the term in Eq.~\eqref{delta_mixed}, and Martina Cottini, Simon Holz, and Yannick Ulrich for sharing their results using dimensional regularization.
Financial support by the SNSF (Project Nos.\ 200020\_175791, PCEFP2\_181117, and  PZ00P2\_174228) and by the Ram\'on y Cajal program (RYC2019-027605-I) of the Spanish MINECO is gratefully acknowledged. 

\appendix

\section{Loop functions}
\label{app:loop_functions}

\subsection{Pole--pole}
\label{sec:loop_pole_pole}

For the pole--pole contribution we need the loop functions for $s'=s''=\lambda^2$:
\begin{align}
 C_0\big(t,\lambda^2\big)&=\frac{1}{\mpi^2-t}\bigg[\log\frac{\lambda^2}{\mpi^2}\log\frac{\mpi^2-t}{\mpi m_e}+\log^2\frac{m_e}{\mpi}-\log^2\frac{\mpi^2-t}{\mpi^2}-\Li_2\bigg(\frac{t}{\mpi^2}\bigg)\bigg],\notag\\
 C_0\big(s,\lambda^2,\lambda^2,\mpi^2\big)&=\frac{1}{s\beta}\bigg[\frac{\pi^2}{6}+\frac{1}{2}\log^2\frac{1-\beta}{1+\beta}+2\Li_2\bigg(\frac{\beta-1}{1+\beta}\bigg)+i\pi \log\frac{1-\beta}{1+\beta}\bigg],\notag\\
 C_0\big(s,\lambda^2,\lambda^2,m_e^2\big)&= \frac{\pi^2+3\log^2\frac{m_e^2}{s}+6i\pi\log\frac{m_e^2}{s}}{6s},\notag\\
 D_0\big(s,t,\lambda^2,\lambda^2\big)&=\frac{2}{s}\bigg(\log\frac{\lambda^2}{s}+i\pi\bigg)\frac{\log\frac{\mpi^2-t}{m_e\mpi}}{\mpi^2-t},
\end{align}
where we suppressed the other arguments  and only kept $m_e$ to regularize collinear singularities at intermediate steps of the calculation. 

\subsection{Pole--dispersive}
\label{sec:loop_pole_dispersive}

For the case $s'\geq 4\mpi^2$, $s''=\lambda^2$ we need the additional loop functions
\begin{align}
 C_0(t,s')&=-\frac{1}{\mpi^2-t}\bigg[\Li_2\bigg(\frac{t}{\mpi^2}\bigg)+\Li_2\bigg(1-\frac{s'(1-\beta')}{2(\mpi^2-t)}\bigg)+\Li_2\bigg(1-\frac{s'(1+\beta')}{2(\mpi^2-t)}\bigg)\notag\\
 &\quad+\frac{\pi^2}{6}+\frac{1}{2}\log^2(1-\beta')+\frac{1}{2}\log^2(1+\beta')-\log\frac{2\mpi^2}{\mpi^2-t}\log\frac{2(\mpi^2-t)}{s'}\bigg],\notag\\
 C_0^{>}(s,s',\mpi^2)&=\frac{1}{s\beta}\bigg[\Li_2\bigg(\frac{1-\beta}{1-\beta'}\bigg)+\Li_2\bigg(\frac{1+\beta'}{1+\beta}\bigg)+\Li_2\bigg(\frac{1-\beta}{1+\beta'}\bigg)+\Li_2\bigg(\frac{1-\beta'}{1+\beta}\bigg)\notag\\
 &\quad+\frac{1}{2}\bigg(\log^2\frac{1-\beta}{1+\beta}+\log^2\frac{1+\beta}{1+\beta'}+\log^2\frac{1+\beta}{1-\beta'}+2\log\frac{1+\beta'}{1-\beta'}\log\frac{\beta-\beta'}{\beta+\beta'}\bigg)\notag\\
 &\quad-\frac{\pi^2}{2}+2\Li_2\bigg(\frac{\beta-1}{1+\beta}\bigg)+i\pi \log\frac{1-\beta}{1+\beta}\bigg],\notag\\
  C_0^{>}(s,s',m_e^2)&= \frac{-\frac{\pi^2}{3}+\frac{1}{2}\log^2\frac{s}{s'}-\log\frac{m_e^2}{s'}\log\frac{s-s'}{s'}+\Li_2\big(\frac{s'}{s}\big)+i\pi \log\frac{m_e^2}{s}}{s},\notag\\
 C_0^{<}(s,s',\mpi^2)&=\frac{1}{s\beta}\bigg[-\Li_2\bigg(\frac{1-\beta'}{1-\beta}\bigg)-\Li_2\bigg(\frac{1+\beta}{1+\beta'}\bigg)+\Li_2\bigg(\frac{1-\beta}{1+\beta'}\bigg)+\Li_2\bigg(\frac{1-\beta'}{1+\beta}\bigg)\notag\\
 &\quad+\frac{1}{2}\bigg(\log^2\frac{1-\beta}{1+\beta}-\log^2\frac{1-\beta}{1-\beta'}+\log^2\frac{1+\beta}{1-\beta'}+2\log\frac{1+\beta'}{1-\beta'}\log\frac{\beta'-\beta}{\beta+\beta'}\bigg)\notag\\
 &\quad+\frac{\pi^2}{6}+2\Li_2\bigg(\frac{\beta-1}{1+\beta}\bigg) 
 \bigg],\notag\\
 C_0^{<}(s,s',m_e^2)&= \frac{-\log\frac{m_e^2}{s'}\log\frac{s'-s}{s'}-\Li_2\big(\frac{s}{s'}\big)}{s},\notag\\
 D_0\big(s,t,s',\lambda^2\big)&=-\frac{\log\frac{\mpi^2-t}{m_e\mpi}\Big[\log\frac{\lambda^2}{\mpi^2}-\log\frac{m_e}{\mpi}+\log\frac{s'^2}{(s'-s)^2}-\log\frac{\mpi^2-t}{\mpi^2}+2\pi i\theta\big(s-s'\big)\Big]}{(\mpi^2-t)(s'-s)}\notag\\
 &-\frac{1}{(\mpi^2-t)(s'-s)}\bigg[\frac{\pi^2}{6}+\Li_2\bigg(1-\frac{s'(1-\beta')}{2(\mpi^2-t)}\bigg)+\Li_2\bigg(1-\frac{s'(1+\beta')}{2(\mpi^2-t)}\bigg)\notag\\
 &\quad+\frac{1}{2}\log^2(1-\beta')+\frac{1}{2}\log^2(1+\beta')-\log\frac{2\mpi^2}{\mpi^2-t}\log\frac{2(\mpi^2-t)}{s'}\bigg],
\end{align}
where $\beta'=\sqrt{1-4\mpi^2/s'}$ and $\gtrless$ indicates that the expression applies to $s\gtrless s'$.

\subsection{Dispersive--dispersive}
\label{sec:loop_dispersive_dispersive}

For general $s',s''$ we write the additional loop functions in the form
\begin{align}
 C_0(s,s',s'',\mpi^2)&=\frac{1}{\pi}\int_{(\sqrt{s'}+\sqrt{s''})^2}^\infty dx \frac{\Im C_0(x,s',s'',\mpi^2)}{x-s},\\
 \Im C_0(s,s',s'',\mpi^2)&=\frac{\pi\theta\big(s-\big(\sqrt{s'}+\sqrt{s''}\big)^2\big)}{s \beta}\log\frac{s-s'-s''-\beta\lambda^{1/2}_s}
 {s-s'-s''+\beta\lambda^{1/2}_s},\notag\\
 D_0\big(s,t,s',s''\big)&=\frac{1}{\pi}\int_{\mpi^2}^\infty dx \frac{\Imt D_0(s,x,s',s'')}{x-t}=\frac{1}{\pi}\int_{(\sqrt{s'}+\sqrt{s''})^2}^\infty dx \frac{\Ims D_0(x,t,s',s'')}{x-s},\notag\\
 \Imt D_0^>(s,t,s',s'')&=\frac{\pi\theta\big(t-\mpi^2\big)}{\sqrt{\Delta(s,t,s',s'')}}
       \bigg[2\pi i \theta\big(s-(\sqrt{s'}+\sqrt{s''})^2\big)\notag\\
       &+\log\frac{(s'+s''-s)(\mpi^2-t)^2+2t s's''+(t-\mpi^2)\sqrt{\Delta(s,t,s',s'')}}{(s'+s''-s)(\mpi^2-t)^2+2t s's''-(t-\mpi^2)\sqrt{\Delta(s,t,s',s'')}}\bigg],\notag\\
 \Imt D_0^<(s,t,s',s'')&=\frac{2\pi\theta\big(t-\mpi^2\big)}{\sqrt{-\Delta(s,t,s',s'')}}
       \bigg[\arctan\frac{(t-\mpi^2)\sqrt{-\Delta(s,t,s',s'')}}{(s'+s''-s)(\mpi^2-t)^2+2ts's''}\notag\\
       &\quad+\pi\theta\Big(\big(s-s'-s''\big)\big(\mpi^2-t\big)^2-2ts' s''\Big)\bigg],\notag\\
                       \Ims D_0(s,t,s',s'')&=\frac{\pi\theta\big(s-(\sqrt{s'}+\sqrt{s''})^2\big)}{\sqrt{\Delta(s,t,s',s'')}}\log\frac{2s s' s''+\lambda_s(\mpi^2-t)+\sqrt{\lambda_s \Delta(s,t,s',s'')}}{2s s' s''+\lambda_s(\mpi^2-t)-\sqrt{\lambda_s \Delta(s,t,s',s'')}},\notag
\end{align}
where 
\beq
\Delta(s,t,s',s'')=\lambda_s\big(\mpi^2-t\big)^2-4ts s's'',\qquad \lambda_s=\lambda(s,s',s''),
\eeq
$\lambda(a,b,c)=a^2+b^2+c^2-2(a b+a c+b c)$, and the expressions are valid in the kinematic region of interest ($s'\geq 4\mpi^2$, $s''\geq 4\mpi^2$, $s\geq 4\mpi^2$, $t\leq 0$, $m_e= 0$). $\gtrless$ indicates that the expression applies for $\Delta(s,t,s',s'')\gtrless0$.


\bibliographystyle{apsrev4-1_mod}
\bibliography{ref}
	
\end{document}